\pdfoutput=1

\documentclass[sigconf]{acmart}
\usepackage{tabularx}
\usepackage{multirow}
\usepackage{csvsimple}
\usepackage{mhchem}
\usepackage{graphicx}
\usepackage{subcaption}
\usepackage{mwe}
\usepackage{float}
\usepackage{placeins}
\usepackage{amsfonts}
\usepackage{booktabs}
\usepackage{siunitx}
\usepackage{array}
\usepackage{hyperref}
\usepackage{pdfpages}
\usepackage{algorithm}
\usepackage{algorithmicx}
\usepackage{algpseudocode}
\usepackage{balance}
\AtBeginDocument{%
  \providecommand\BibTeX{{%
    \normalfont B\kern-0.5em{\scshape i\kern-0.25em b}\kern-0.8em\TeX}}}

\setcopyright{acmcopyright}
\copyrightyear{2020} 
\acmYear{2020} 
\setcopyright{acmlicensed}\acmConference[e-Energy'20]{The Eleventh ACM International Conference on Future Energy Systems}{June 22--26, 2020}{Virtual Event, Australia}
\acmBooktitle{The Eleventh ACM International Conference on Future Energy Systems (e-Energy'20), June 22--26, 2020, Virtual Event, Australia}
\acmPrice{15.00}
\acmDOI{10.1145/3396851.3397682}
\acmISBN{978-1-4503-8009-6/20/06}

\begin{document}
\emergencystretch 3em
\title[Electricity market validation using genetic algorithms]{Long-term electricity market agent based model validation using genetic algorithm based optimization}


\author{Alexander J. M. Kell}
\affiliation{%
  \department{School of Computing}
  \institution{Newcastle University}
  \city{Newcastle upon Tyne}
  \country{UK}
}
\email{a.kell2@newcastle.ac.uk}

\author{Matthew Forshaw}
\affiliation{%
  \department{School of Computing}
  \institution{Newcastle University}
  \city{Newcastle upon Tyne}
  \country{UK}
}
\email{matthew.forshaw@newcastle.ac.uk}

\author{A. Stephen McGough}
\affiliation{%
  \department{School of Computing}
  \institution{Newcastle University}
  \city{Newcastle upon Tyne}
  \country{UK}
}
\email{stephen.mcgough@newcastle.ac.uk}
%


\renewcommand{\shortauthors}{Kell et al.}

\begin{abstract}
 
Electricity market modelling is often used by governments, industry and agencies to explore the development of scenarios over differing timeframes. For example, how would the reduction in cost of renewable energy impact investments in gas power plants or what would be an optimum strategy for carbon tax or subsidies?  

Cost optimization based solutions are the dominant approach for understanding different long-term energy scenarios. However, these types of models have certain limitations such as the need to be interpreted in a normative manner, and the assumption that the electricity market remains in equilibrium throughout. Through this work, we show that agent-based models are a viable technique to simulate decentralised electricity markets. The aim of this paper is to validate an agent-based modelling framework to increase confidence in its ability to be used in policy and decision making. 

Our framework can model heterogeneous agents with imperfect information. The model uses a rules-based approach to approximate the underlying dynamics of a real world, decentralised electricity market. We use the UK as a case study, however, our framework is generalisable to other countries. We increase the temporal granularity of the model by selecting representative days of electricity demand and weather using a $k$-means clustering approach. 
 
We show that our framework can model the transition from coal to gas observed in the UK between 2013 and 2018. We are also able to simulate a future scenario to 2035 which is similar to the UK Government, Department for Business and Industrial Strategy (BEIS) projections. We show a more realistic increase in nuclear power over this time period. This is due to the fact that with current nuclear technology, electricity is generated almost instantaneously and has a low short-run marginal cost \cite{Department2016}.

\end{abstract}

%
%


\maketitle

\section{Introduction}


Impacts on natural and human systems due to climate change have already been observed, with many land and ocean ecosystems having changed. A rise in carbon emissions increases the risk of severe impacts on the world, such as rising sea levels and heat waves \cite{Masson-Delmotte2018}. A study by Cook \textit{et al.} demonstrated that 97\% of scientific literature concurred that recent global warming was anthropogenic \cite{Cook2013}, thus limiting global warming requires limiting the total cumulative global anthropogenic emissions of  \ce{CO2} \cite{Masson-Delmotte2018}. 

Global carbon emissions from fossil fuels have increased since 1900 \cite{boden2017global}. Fossil-fuel based electricity generation sources such as coal and natural gas currently provide 65\% of global electricity. Low-carbon sources such as solar, wind and nuclear provide 35\% \cite{BP2018}. To halt this increase in \ce{CO2} emissions, a transition of the energy system towards a renewable energy system is required. Such a transition needs to be performed in a gradual and non-disruptive manner. This ensures that there are no electricity shortages that would cause damage to businesses, consumers and the economy. 

To ensure such a transition, energy modelling is often used by governments, industry and agencies to explore possible scenarios under different variants of government policy, future electricity generation costs and energy demand. These energy modelling tools aim to mimic the behaviour of energy systems through different sets of equations and data sets to determine the energy interactions between different actors and the economy \cite{Machado2019}.


Optimization based solutions are the dominant approach for understanding long-term energy policy \cite{Chappin2017}. However, the results of these models should be interpreted in a normative manner. For example, how investment and policy choices should be carried out, under certain assumptions and scenarios. The processes which emerge from an equilibrium model remain a black-box, making it difficult to fully understand the underlying dynamics of the model \cite{Chappin2017}.

In addition to this, optimization models do not allow for endogenous behaviour to emerge from typical market movements, such as investment cycles \cite{Chappin2017, Gross2007}. By modelling these naturally occurring behaviours, policy can be designed that is robust against movements away from the optimum/equilibrium. Thus, helping policy to become more effective in the real world.


The work presented in this paper builds on the agent-based model (ABM), ElecSim, developed by Kell \textit{et al.} \cite{Kell}. ABMs differ from optimization models by the fact that they can explore `\textit{what-if}' questions regarding how a sector could develop under different prospective policies, as opposed to determining optimal trajectories. ABMs are particularly pertinent in decentralized electricity markets, where a centralized actor does not dictate investments made within the electricity sector. ABMs can closely mimic the real world by, for example, modelling irrational agents, in this case, Generation Companies (GenCos) with incomplete information in uncertain situations \cite{Ghorbani2014}. 

There is a desire to validate the ability of energy-models to make long-term predictions. Validation increases confidence in the outputs of a model and leads to an increase in trust amongst both the public and policymakers. Energy models, however, are frequently criticized for being insufficiently validated, with the performance of models rarely checked against historical outcomes \cite{Beckman2011}.

In answer to this, we postulate that ABMs can provide accurate information to decision-makers in the context of electricity markets. We increase the temporal granularity of our prior work \cite{Kell} and use genetic algorithms (GAs) to tune the model to observed data enabling us to perform validation. This enables us to understand the parameters required to observe certain phenomena, as well as use these fitted parameters to make inferences about the future.


We use a GA approach to find an optimal set of price curves predicted by generation companies (GenCos) that adequately model observed investment behaviour in the real-life electricity market in the United Kingdom. Similar techniques can be employed for other countries of various sizes \cite{Kell}.

 Similarly to Nahmmacher \textit{et al.} we demonstrate how clustering of multiple relevant time series such as electricity demand, solar irradiance and wind speed can reduce computational time by selecting representative days ~\cite{Nahmmacher2016}. In this context, representative days are a subset of days that have been chosen due to their ability to approximate the weather and electricity demand in an entire year. Distinct to Nahmacher \textit{et al.} we use a $k$-means clustering approach \cite{forgy65} as opposed to a hierarchical clustering algorithm described by Ward \cite{doi:10.1080/01621459.1963.10500845}. We chose the $k$-means clustering approach due to the previous success of this technique in clustering time series \cite{Kell2018a}. 

We measure the accuracy of projections for our improved ABM with those of the UK Government's Department for Business, Energy and Industrial Strategy (BEIS) for the UK electricity market between 2013 and 2018. In addition to this, we compare our projections from 2018 to 2035 to those made by BEIS in 2018 \cite{DBEIS2019}.

We can model the transitional dynamics of the electricity mix in the UK between 2013 and 2018. During this time there was an ${\sim}88\%$ drop in coal use, ${\sim}44\%$ increase in Combined Cycle Gas Turbines (CCGT), ${\sim}111\% $ increase in wind energy and increase in solar from near zero to ${\sim}1250$MW. We are therefore able to test our model in a transition of sufficient magnitude.


We show in this paper that ABMs can adequately mimic the behaviour of the UK electricity market under the same specific scenario conditions. Concretely, we show that under an observed carbon tax strategy, fuel price and electricity demand, the model ElecSim closely matches the observed electricity mix between 2013 and 2018. We achieve this by determining an exogenous predicted price duration curve using a GA to minimize the error between observed and simulated electricity mix in 2018. The predicted price curve is an arrangement of all price levels in descending order of magnitude. The predicted price duration curve achieved is similar to that of the simulated price duration curve in 2018, increasing confidence in the underlying dynamics of our model. The model parameters and data are instantiated with data on fuel costs, carbon tax prices and historical generator costs going back to the years 1990, 2009 and 1981, respectively.

In addition, we compare our projections to those of the BEIS reference scenario from 2018 to 2035~\cite{DBEIS2019}. To achieve this, we use the same GA optimization technique as during our validation stage, optimizing for predicted price duration curves. Our model demonstrates that we can closely match the projections of BEIS by finding a set of realistic price duration curves which are subject to investment cycles. Our model, however, exhibits a more realistic step-change in nuclear output than that of BEIS. This is because, while BEIS projects a gradual increase in nuclear output, our model projects that nuclear output will grow instantaneously at a single point in time as a new nuclear power plant comes online. 

This allows us to verify the scenarios of other models, in this case, BEIS' reference scenario, by ascertaining whether the optimal parameters required to achieve such scenarios are realistic. In addition to this, we can use these parameters to analyze `\textit{what-if}' questions with further accuracy.



We increased the temporal granularity of the model using a $k$-means clustering approach to select a subset of representative days for wind speed, solar irradiance and electricity demand. This subset of representative days enabled us to approximate an entire year and only required a fraction of the total time-steps that would be necessary to model each day of a year. This enabled us to decrease execution time by ${\sim}40\times$. We show that we can provide an accurate framework, through this addition, to allow policy and decision-makers, as well as the public, explore the effects of policy on investment in electricity generators.

We demonstrate that with a GA approach, we can optimize parameters to improve the accuracy of our model. Namely, we optimize the predicted electricity price, the uncertainty of this electricity price and nuclear subsidy. We use validation to verify our model using the observed electricity mix between 2013-2018.

The main contribution of this work is to demonstrate that it is possible for ABMs to accurately model transitions in the UK electricity market. This was achieved by comparing our simulated electricity mix to the observed electricity mix between 2013 and 2018. In this time, a transition from coal to natural gas was observed. We demonstrate that a high temporal granularity is required to accurately model fluctuations in wind and solar irradiance for intermittent renewable energy sources.


In Section \ref{lit-review} we introduce a review of techniques used for validating electricity market models as well as fundamental challenges of electricity model validation. Section \ref{ssec:prob_formulation} explores our approach to validate our model. In Section \ref{sec:details} we discuss the modifications made to our model to improve the results of validation. In Section \ref{sec:results} we present our results, and we conclude in Section \ref{sec:conclusion}.

\section{Literature Review}
\label{lit-review}



In this section, we cover the difficulties inherent in validating energy models and the approaches taken in the literature.

\subsection{Limits of Validating Energy Models}

Beckman \textit{et al.} state that questions frequently arise as to how much faith one can put in energy model results. This is due to the fact that the performance of these models as a whole are rarely checked against historical outcomes~\cite{Beckman2011}.

Under the definition by Hodges \textit{et al.} \cite{Hodges} long-range energy forecasts are not validatable \cite{Craig2002}. Under this definition, validatable models must be observable, exhibit constancy of structure in time, exhibit constancy across variations in conditions not specified in the model and it must be possible to collect ample data \cite{Hodges}.

While it is possible to collect data for energy models, the data covering important characteristics of energy markets are not always measured. Furthermore, the behaviour of the human population and innovation are neither constant nor entirely predictable. This leads to the fact that static models cannot keep pace with long-term global evolution. Assumptions made by the modeller may be challenged in the form of unpredictable events, such as the oil shock of 1973 \cite{Craig2002}.

This, however, does not mean that energy-modelling is not useful for providing advice in the present. A model may fail at predicting the long-term future because it has forecast an undesirable event, which led to a pre-emptive change in human behaviour—thus avoiding the original scenario that was predicted. This could, therefore, be viewed as a success of the model.

Schurr \textit{et al.} argued against predicting too far ahead in energy modelling due to the uncertainties involved \cite{Schurr_1961}. However, they specify that long-term energy forecasting is useful to provide basic information on energy consumption and availability, which is helpful in public debate and in guiding policymakers.

Ascher concurs with this view and states that the most significant factor in model accuracy is the time horizon of the forecast; the more distant the forecast target, the less accurate the model. This can be due to unforeseen changes in society as a whole ~\cite{gillespie_1979}.

It is for these reasons that we focus on a shorter-term (5-year) horizon window when validating our model. This enables us to have increased confidence that the dynamics of the model work without external shocks and can provide descriptive advice to stakeholders. However, it must be noted that the UK electricity market exhibited a fundamental transition from coal to natural gas electricity generation during this period, meaning that a simple data-driven modelling approach would not work.

In addition to this short-term cross-validation, we compare our long-term projections to those of BEIS from 2018 to 2035. It is possible that our projections and those of BEIS could be wrong, however, this allows us to test a particular scenario with different modelling approaches thoroughly, and allow for the possibility to identify potential flaws in the models.

\subsection{Validation Examples}

In this section, we explore a variety of approaches used in the literature for energy model validation.

The model OSeMOSYS \cite{Howells2011} is validated against the similar model MARKAL\slash TIMES through the use of a case study named UTOPIA. UTOPIA is a simple test energy system bundled with ANSWER; a graphical user interface packaged with the MARKAL model generator \cite{Hunter2013, Noble2004}. Hunter \textit{et al.} use the same case study to validate their model Temoa \cite{Hunter2013}. In these cases, MARKAL\slash TIMES is seen as the "gold standard". In this paper, however, we argue that the ultimate gold standard should be real-world observations, as opposed to a hypothetical scenario.

The model PowerACE demonstrates that their modelling approach achieves realistic prices, however, they do not indicate success in modelling GenCo investment over a prolonged time period \cite{Ringler2012}.

Barazza \textit{et al}. validate their model, BRAIN-Energy, by comparing their results with a few years of historical data. However, they do not compare the simulated and observed electricity mix \cite{Barazza2020}.

Koomey \textit{et al.} expresses the importance of conducting retrospective studies to help improve models \cite{Koomey2003}. In this case, a model can be rerun using historical data in order to determine how much of the error in the original forecast resulted from structural problems in the model itself, or how much of the error was due to incorrect specification of the fundamental drivers of the forecast \cite{Koomey2003}.

A retrospective study published in 2002 by Craig \textit{et al.} focused on the ability of forecasters to predict electricity demand from the 1970s accurately \cite{Craig2002}. They found that actual energy usage in 2000 was at the very lowest end of the forecasts, with only a single exception. They found that these forecasts underestimated unmodelled shocks such as the oil crises which led to an increase in energy efficiency.

Hoffman \textit{et al.} also developed a retrospective validation of a predecessor of the current MARKAL\slash TIMES model, named Reference Energy System \cite{Hoffman_1973}, and the Brookhaven Energy System Optimization Model \cite{ERDA_48}. These were studies applied in the 70s and 80s to develop projections to the year 2000. This study found that the models could be descriptive but was not entirely accurate in terms of predictive ability. They found that emergent behaviours in response to policy had a substantial impact on forecasting accuracy. The study concluded that forecasts must be expressed in highly conditioned terms \cite{Hoffman2011}.

\section{Problem Formulation}
\label{ssec:prob_formulation}

In this section, we detail the approach taken in this paper to validate our model, including the parameters used for optimization. 

Specifically, we use a GA to find the predicted price duration curves, which lead to the smallest error between our simulated electricity mix and the scenarios tested. The scenarios examined here are the observed electricity mix of the UK between 2013 and 2018 and the BEIS reference scenario projected in 2018 till 2035. We also optimize for nuclear subsidy and uncertainty in the price duration curves.

\subsection{Optimization Variables}

For GenCos to adequately make investments, they must formulate an expectation of future electricity prices over the lifetime of a plant. For this paper, we use the net present value (NPV) metric to compare investments.

NPV provides a method for evaluating and comparing investments with cash flows spread over many years, making it suited for evaluating power plants which have a long lifetime.  

Equation \ref{eq:npv_eq} is the calculation of NPV, where $t$ is the year of the cash flow, $i$ is the discount rate, $N$ is the total number of years or lifetime of the power plant, and $R_t$ is the net cash flow of the year $t$.

The total net cash flow, $R_t$, is calculated by subtracting both the operational and capital costs from revenue over the expected lifetime of the prospective plant per year $t$. The revenue gained by each prospective plant is the expected price they will gain per expected quantity of MWh sold over the expected lifetime of the plant. This is shown formally in Equation \ref{eq:revenue_total}:

\begin{equation} \label{eq:npv_eq}
NPV(t, N) = \sum_{t=0}^{N}\frac{R_t}{(1+i)^t}
\end{equation}

\begin{equation}
\label{eq:revenue_total}
    R_t = 
    \sum\limits_{t=0}^T 
    \sum\limits_{h=0}^H
    \sum\limits_{m=0}^M \left(
    m_{h,t}(PPDC_{h,t}
    -
    C_{var_{h,t}})\right)
    - C_c
\end{equation}

\noindent where $m_{h,t}$ is the expected quantity of megawatts sold in hour $h$ of year $t$. $M$ is the total number of megawatts sold over the respective time period. $PPDC_{h,t}$ is the predicted price duration curve at year $t$ and hour $h$. $C_{var_{h,t}}$ is the variable cost of the power plant, which is dependent on expected megawatts of electricity produced, $C_c$ is the capital cost. Capital costs are included due to the fact that these are a large expense and thus should be subtracted from the total net cash flow.

The predicted price duration curve ($PPDC_{h,t}$) is an expectation of future electricity prices over the lifetime of the plant. The $PPDC_{h,t}$ is a function of supply and demand. However, with renewable electricity generator costs falling, future prices are uncertain and largely dependent upon long-term scenarios of electricity generator costs, fuel prices, carbon taxes and investment decisions. \cite{IRENA2014}. Due to the uncertainty of future electricity prices over the horizon of the lifetime of a power plant, we have set future electricity prices as an exogenous variable that can be set by the user in ElecSim.

To gain an understanding of expected electricity prices that lead to particular scenarios, we use a GA optimization approach. This enables us to understand the range of future electricity prices that lead to specific scenarios developing. In addition, it allows us to understand whether the parameters required for specific scenarios to develop are realistic. This enables us to check the assumptions of our model and the likelihood of scenarios.

Using these optimized parameters, we are better able to further explore `\textit{what-if}' scenarios.

\subsection{Validation with Observed Data}
\label{sssec:validation}

To verify the accuracy of the underlying dynamics of ElecSim, the model was initialized to data available in 2013 and allowed to develop until 2018. We used a GA to find the optimum price duration curve predicted ($PPDC$) by the GenCos ten years ahead of the year of the simulation. This $PPDC$ was used to model expected rate of return of prospective generation types, as shown in Equations \ref{eq:npv_eq} and \ref{eq:revenue_total}. 

The GA's objective was to reduce the error of simulated and observed electricity mix in the year 2018 by finding a suitable $PPDC$ used by each of the GenCos for investment evaluation.

\subsubsection{Scenario}

For this experiment, we initialized ElecSim with parameters known in 2013 for the UK. ElecSim was initialized with every power plant and respective GenCo that was in operation in 2013 using the BEIS DUKES dataset \cite{dukes_511}. The funds available to each of the GenCos was taken from publicly released official company accounts at the end of 2012 \cite{companies_house}.

To ensure that the development of the electricity market from 2013 to 2018 was representative of the actual scenario between these years, we set the exogenous variables, such as carbon and fuel prices, to those that were observed during this time period. 

The data for the observed EU Emission Trading Scheme (ETS) price between 2013 and 2018 was taken from \cite{eu-ets}. Fuel prices for each of the fuels were taken from \cite{beis_fuel_price}. The electricity load data was modelled using data from \cite{gridwatch}, offshore, and onshore wind and solar irradiance data was taken from \cite{Pfenninger2016}. There were three known significant coal plant retirements in 2016. These were removed from the simulation at the beginning of 2016.

\subsubsection{Optimisation problem}
\label{ssec:optimisation-problem}

The price duration curve was modelled linearly in the form $y=mx+c$, where $y$ is the cost of electricity, $m$ is the gradient, $x$ is the demand of the price duration curve, and $c$ is the intercept. This is a limiting factor. However, we propose that the linear model sufficiently models the price duration curve, while maintaining a model with only two parameters. 

Equation \ref{eq:problem_formulation} details the optimisation problem formally:

\begin{equation}
    \label{eq:problem_formulation}
    \min_{m,c} \sum\limits_{o\in O}\left(
    \frac{\left|A_o-f_o(m,c)\right|}
    {\left|\left|O\right|\right|}
    \right)
\end{equation}

\noindent where $o\in O$ refers to the average percentage electricity mix during 2018 for wind (both offshore and onshore generation), nuclear, solar, CCGT, and coal, where $O$ refers to the set of these values. $A_o$ refers to observed electricity mix percentage for the respective generation type in 2018. $f_o(m,c)$ refers to the simulated electricity mix percentage for the respective generation type, also in 2018. The input parameters to the simulation are $m$ and $c$ from the linear $PPDC$, previously discussed, ie. $y=mx+c$. $\left|\left|O\right|\right|$ refers to the cardinality of the set.

\subsection{Long-Term Scenario Analysis}
\label{sssec:scen-analysis}

In addition to verifying the ability for ElecSim to mimic observed investment behaviour over five years, we compared ElecSim's long-term behaviour to that of the UK Government's Department for Business, Energy and Industrial Strategy (BEIS) \cite{DBEIS2019}. This scenario shows the projections of generation by technology for all power producers from 2018 to 2035 for the BEIS reference scenario. This is the same scenario as discussed in the next section, \ref{sssec:scenario-details}.

\subsubsection{Scenario}
\label{sssec:scenario-details}

We initialized the model to 2018 based on \cite{Kell}. The scenario for the development of fuel prices and carbon prices were matched to that of the BEIS reference scenario \cite{DBEIS2019}.

\subsubsection{Optimisation problem} The optimization approach taken was a similar process to that discussed in Sub-Section \ref{ssec:optimisation-problem}, namely using a GA to find the optimum expected price duration curve. However, instead of using a single expected price duration curve for each of the agents for the entire simulation, we used a different expected price duration curve for each year, leading to 17 different curves. This enabled us to model the non-static dynamics of the electricity market over this extended time period. 

In addition to optimizing for multiple expected price duration curves, we optimized for a nuclear subsidy, $S_n$. Further, we optimized for the uncertainty in the expected price parameters $m$ and $c$, named $\sigma_m$ and $\sigma_c$ respectively, where $\sigma$ is the standard deviation in a normal distribution. $m$ and $c$ are the parameters for the predicted price duration curve, as previously defined, of the form $y=mx+c$. This enabled us to model the different expectations of future price curves between the independent GenCos. The addition of a nuclear subsidy as a parameter is due to the likely requirement for Government to provide subsidies for new nuclear \cite{Suna2016}.

A modification was made to the reward algorithm for the long-term scenario case. Rather than using the discrepancy between observed and simulated electricity mix in the final year (2018) as the reward, a summation of the error metric for each simulated year was used. This is detailed formally in Equation \ref{eq:long-term-reward}:

\begin{equation}
    \label{eq:long-term-reward}
    \min_{m_y\in M,c_y\in C} 
    \sum\limits_{y\in Y}
    \sum\limits_{o\in O}\left(
    \frac{\left|A_{y_o}-f_{y_o}(m_y,c_y)\right|}
    {\left|\left|O\right|\right|}
    \right)
\end{equation}

\noindent where $M$ and $C$ are the sets of the 17 parameters of $m_y$ and $c_y$ respectively for each year, $y$. $y\in Y$ refers to each year between 2018 and 2035. $m_y$ and $c_y$ refer to the parameters for the predicted price duration curve, of the form $y=mx+c$ for the year $y$. $A_{y_o}$ refers to the actual electricity mix percentage for the year $y$ and generation type $o$. Finally, $f_{y_o}(m_y,c_y)$ refers to the simulated electricity mix percentage with the input parameters to the simulation of $m$ and $c$ for the year $y$.

\section{Implementation details}
\label{sec:details}

In this section, we discuss the changes made to Kell \textit{et al.} to improve the results of validation \cite{Kell}. Further, we introduce the GA used to find the optimal parameter sets.

ElecSim is made up of six distinct sections: 1) power plant data; 2) scenario data; 3) the time-steps of the algorithm; 4) the power exchange; 5) the investment algorithm and 6) the generation companies (GenCos) as agents. The model is given data on fuel costs, carbon tax prices and historical generator costs going back to the years 1990, 2009 and 1981 respectively. ElecSim has been previously published \cite{Kell}. However, we have made amendments to the original work in the form of efficiency improvements to decrease compute time as well as increase the granularity of time-steps from yearly to representative days. Representative days, in this context, are a subset of days which, when scaled up to 365 days can adequately represent a year. 

In this paper, we initialized the model to a scenario of the United Kingdom as an example. However, the fundamental dynamics of the model remain the same for other decentralized electricity markets. In this section, we detail the modifications we made to ElecSim for this paper. Further details of the design decisions of ElecSim are discussed in \cite{Kell}.


\subsection{Representative Days}
\label{ssec:representative_days}

In previously published work, ElecSim modelled a single year as 20 time-steps for solar irradiance, onshore and offshore wind and electricity demand~\cite{Kell}. Similarly to findings of other authors, this relatively low number of time-steps led to an overestimation of the uptake of intermittent renewable energy resources (IRES) and an underestimation of flexible technologies~\cite{Ludig2011,Haydt2011}. This is because the full intermittent nature of renewable energy could not be accurately modelled in such a small number of time-steps.


To address this problem, while maintaining a tractable execution time, we approximated a single year as a subset of proportionally weighted, representative days. This enabled us to reduce computation time. Each representative day consisted of 24 equally separated time-steps, which model hours in a day. Hourly data was chosen, as this was the highest resolution of the dataset available for offshore and onshore wind and solar irradiance \cite{Pfenninger2016}. A lower resolution would allow us to model more days. However, we would lose accuracy in terms of the variability of the renewable energy sources.

Similarly to Nahmmacher \textit{et al.} we used a clustering technique to split similar days of weather and electricity demand into separate groups. We then selected the historic day that was closest to the centre of the cluster, known as the medoid, as well as the average of the centre, known as the centroid ~\cite{Nahmmacher2016}. Distinct to Nahmmacher, however, we used the $k$-means clustering algorithm~\cite{forgy65} as opposed to the Ward's clustering algorithm \cite{doi:10.1080/01621459.1963.10500845}. This was due to the ability for the $k$-means algorithm to cluster time-series into relevant groups \cite{Kell2018a}. These days were scaled proportionally to the number of days within their respective cluster to approximate a total of 365 days.

Equation \ref{eq:medoids_series} shows the series for a medoid or centroid, selected by the $k$-means algorithm:

\begin{equation}
\label{eq:medoids_series}
    P^{x,i}_{h}=\{P_1, P_2, \ldots, P_{24}\}
\end{equation}

\noindent where $P^{x,i}_{h}$ is the medoid for series $x$, where $x\in X$ refers to offshore wind capacity factor, onshore wind capacity factor, solar capacity factor and electricity demand, $h$ is the hour of the day and $i$ is the respective cluster. $\{P_1, P_2, \ldots , P_{24}\}$ refers to the capacity values at each hour of the representative day.

We then calculated the weight of each cluster. This gave us a method of assigning the relative importance of each representative day when scaling the representative days up to a year. The weight is calculated by the proportion of days in each cluster. This gives us a method of determining how many days within a year are similar to the selected medoid or centroid. The calculation for the weight of each cluster is shown by Equation \ref{eq:cluster_weight}:

\begin{equation}
\label{eq:cluster_weight}
    w_i = \frac{n_i}{||N||} 
\end{equation} 

\noindent where $w_i$ is the weight of cluster $i$, $n_i$ is the number of days in cluster $i$, and $||N||$ is the total number of days that have been used for clustering. In this case, $||N||$ is equal to 2772 days (data used between 27-05-2011 and 27-12-2018).

The next step was to scale up the representative days to represent the duration curve of a full year. We achieved this by using the weight of each cluster, $w_i$, to increase the number of hours that each capacity factor contributed in a full year. Equation \ref{eq:scaled-medoid} details the scaling process to turn the medoid or centroid, shown in Equation \ref{eq:medoids_series}, into a scaled day. Where $\widetilde{P}^{x,i}_{h}$ is the scaled day:

\begin{equation}
\label{eq:scaled-medoid}
    \widetilde{P}^{x,i}_{h} =  \{P_{1w_i}, P_{2w_i}, \ldots, P_{24w_i}\}
\end{equation} 

\noindent Equation \ref{eq:scaled-medoid} effectively extends the length of the day proportional to the number of days in the respective cluster. For example, for $P_{24w_i}$, the 24th hour of the day is scaled by its respective weight $w_i$ found by Equation \ref{eq:cluster_weight}.


 Finally, each of the scaled representative days were concatenated to create the series used for the calculations which required the capacity factors and the respective number of hours that each capacity factor contributed to the year. Equation \ref{eq:total_time_series} displays the total time series for series $x$, where each scaled medoid is concatenated to produce an approximated time series, $\widetilde{P}^x$:

\begin{equation}
\label{eq:total_time_series}
    \widetilde{P}^x=\left(\widetilde{P}^{x,1}_{h},\widetilde{P}^{x,2}_{h},\ldots, \widetilde{P}^{x,||N||}_{h}\right)
\end{equation}

\noindent the total number of hours in the approximated time series, $\widetilde{P}^x$, is equal to the number of hours in a day multiplied by the number of days in a year, which gives the total number of hours in a year ($24\times 365=8760$), as shown by Equation \ref{eq:total_scale}:

\begin{equation}
\label{eq:total_scale}
    \sum\limits_{w\in W}\sum\limits_{t=1}^{T=24}\left(w_i t\right)=24\times 365=8760
\end{equation}

\noindent where $w\in W$ is the set of clusters. Effectively, Equation \ref{eq:total_time_series} is a time series of an entire year, using only a subset of representative days, scaled proportionally to the amount of time series within each cluster of the k-means algorithm.

\subsection{Error Metrics}

To measure the validity of our approximation using representative days and also compare the optimum number of days, or clusters, we used a technique similar to Poncelet \textit{et al.} \cite{Dhaeseleer2015, Poncelet2017}. We trialled the number of clusters against three different metrics: correlation ($CE_{av}$), normalised root mean squared error ($nRMSE$) and relative energy error ($REE_{av}$). We used the duration curve here as opposed to the actual and approximated time series as ElecSim uses a day-ahead market. Therefore the order in which the market is cleared is not of importance.

$CE_{av}$ details the correlation between the different time-series: solar irradiance, onshore and offshore wind speed and electricity demand. These time-series are inextricably linked, due to the effect of heat generated by solar on the wind, and the effect of weather on electricity demand. $nRMSE$ is the error between each time-series. $REE_{av}$ is the average value of the respective time series, which is distinct to the overall error between each time-series, over a more extended time period, as monitored by $nRMSE$.

$REE_{av}$ is the average value over all the considered time series $\widetilde{P}^x{\in} \widetilde{P}$ compared to the observed average value of the set $P^x\in P$. Where $P^x\in P$ are the observed time series and $\widetilde{P}^x{\in} \widetilde{P}$ are the scaled, approximated time series using representative days. $REE_{av}$ is shown formally by Equation \ref{eq:ree_av}:

\begin{equation}
\label{eq:ree_av}
    REE_{av}=\frac
    {\sum\limits_{P^x{\in} P}\left(\left|
    \frac
    {\sum\limits_{t\in T}DC_{P^x_t}-\sum\limits_{t\in T}\widetilde{DC}_{\widetilde{P}^x_t}}
    {\sum\limits_{t\in T}DC_{P^x_t}}
    \right|\right)
    }
    {\left|\left|P\right|\right|}
\end{equation}

\noindent where $DC_{P^x_t}$ is the duration curve for $P^x$ and $DC_{\widetilde{P}^x_t}$ is the duration curve for $\widetilde{P}^x$. In this context, the duration curve can be constructed by sorting the capacity factor and electrical load data from high to low. The $x-$axis for the DC exhibits the proportion of time that each capacity factor represents. The approximation of the duration curve is represented in this text as $\widetilde{DC}_{\widetilde{p}^x}$. $t\in T$ refers to a specific time step of the original time series. $\widetilde{DC}$ refers to the approximated duration curve for $\widetilde{P}^x$. Note that in this text $\left|\cdot\right|$ refers to the absolute value, and $\left|\left|\cdot\right|\right|$ refers to the cardinality of a set and $\left|\left|P\right|\right|$ refers to the total number of of considered time series.

Specifically, the sum of the observed values, $P^x$, and approximated values, $\widetilde P^x$, for all of the time series are summed. The proportional difference is found, which is summed for each of the different series, $x$, and divided by the number of series, to give $REE_{av}$.



Another requirement is for the distribution of load and capacity factors for the approximated series to correspond to the observed time series. It is crucial that we can account for both high and low levels of demand and capacity factor for IRES generation. This enables us to model for times where flexible generation capacity is required.

The distribution of values can be represented by the duration curve ($DC$) of the time series. Therefore, the average normalised root-mean-square error ($NRMSE_{av}$) between each $DC$ is used as an additional metric, as shown by Equation \ref{eq:nrmse_av}:

\begin{equation}
\label{eq:nrmse_av}
    NRMSE_{av}=\frac
    {\sum\limits_{P^x{\in} P}\left(\frac
    {\sqrt{
    \frac{1}{\left|\left|T\right|\right|}
    \cdot
    \sum\limits_{t\in T}(DC_{P^x_t}-\widetilde{DC}_{\widetilde{P}^x_t})^2}
    }
    {max(DC_{P^x})-min(DC_{P^x})}
    \right)}
    {\left|\left|P\right|\right|}.
\end{equation}

Specifically, the difference between the approximated and observed duration curves for each time-step $t$ is calculated. The average value is then taken of these differences. This average value is then normalized for the respective time series $P^x$. The average of these average normalized values for each time series are then taken to provide a single metric, $NRMSE_{av}$.

The final metric used is the correlation between the different time series. This is used due to the fact that wind and solar output influences the load within a single region, solar and wind output are correlated, as well as offshore and onshore wind levels within the UK. This is referred to as the average correlation error ($CE_{av}$) and shown formally by Equation \ref{eq:ce_av}:

\begin{equation}
\label{eq:ce_av}
    CE_{av}=\frac{2}{\left|\left|P\right|\right|\cdot(\left|\left|P\right|\right|-1)}\cdot
    \left(
    \sum\limits_{p_i\in P}\sum\limits_{p_j\in P,j>i}
    \left|
    corr_{p_i,p_j}-\widetilde{corr}_{p_i,p_j}
    \right|
    \right)
\end{equation}

\noindent where $corr_{p1,p2}$ is the Pearson correlation coefficient between two time series $p_1,p_2\in P$, shown by Equation \ref{eq:corr}. Here, $V_{p1,t}$ represents the value of time series $p_1$ at time step t:

\begin{equation}
\label{eq:corr}
    corr_{p1,p2}=\frac
    {\sum\limits_{t\in T}\left(\left(V_{p1,t}-\overline{V}_{p1}\right)\cdot\left(V_{p2,t}-\overline{V}_{p2}\right)\right)}
    {\sqrt{
    \sum\limits_{t\in T} \left(V_{p_1,t}-\overline{V}_{p1}\right)^2\cdot\sum\limits_{t\in T}\left(V_{p2,t}-\overline{V}_{p2}\right)^2
    }}.
\end{equation}

\subsection{Integrating higher temporal granularity}

To integrate the additional temporal granularity of the model, extra time-steps were taken per year. The higher temporal granularity of the model enabled us to accurately model the hourly fluctuations in solar and wind which leads to more accurate expectations of the investment opportunities of these technologies ~\cite{Ludig2011,Haydt2011}.

\begin{figure}
\centering
\includegraphics[width=0.49\textwidth]{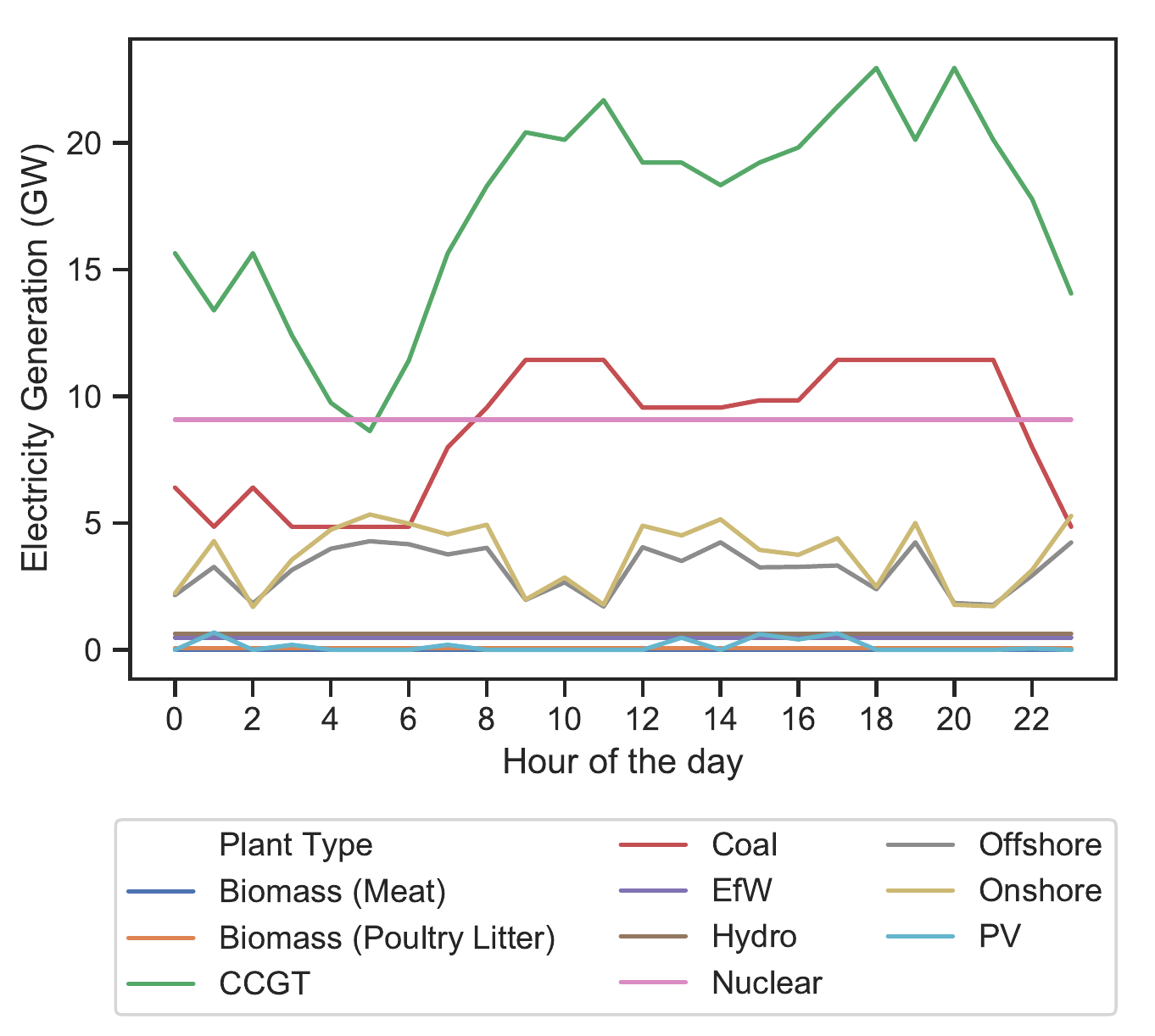}
\caption{Example of a single day of dispatched supply.}
\label{fig:single_dispatched_day}
\end{figure}

GenCos make bids at the beginning of every time-step, and the Power Exchange matches demand with supply in merit-order dispatch using a uniform pricing market. An example of the electricity mix in a single representative day is shown in Figure \ref{fig:single_dispatched_day}. 

%

Figure \ref{fig:single_dispatched_day} displays the high utilization of low marginal-cost generators such as nuclear, wind and photovoltaics. At hour 19, an increase in offshore wind leads to a direct decrease in CCGT. In contrast to this, a decrease in offshore and onshore between the hours of 8 and 12 leads to an increase in dispatch of coal and CCGT. One would expect this behaviour to prevent blackouts and meet demand at all times. This process has enabled us to match fluctuations in IRES more closely.

\subsection{Genetic Algorithms}

Genetic Algorithms (GAs) are a type of evolutionary algorithm which can be used for optimization. We chose the GA for this application due to its ability to find good solutions with a limited number of simulation runs, its ability for parallel computation and its ability to find global optima. These characteristics are useful in our context, as a single simulation can take up to 36 hours. 

In this section, we detail the GA used in this paper. Initially, a population $P_{0}$ is generated for generation 0. This population of individuals is used for the parameters to the simulation. The output of the simulations for each of the individuals are then evaluated. A subset of these individuals $C_{t+1} \subset P_{t}$ are chosen for mating. This subset is selected proportional to their fitness, with `Fitter' individuals having a higher chance of reproducing to create the offspring group $C'_{t+1}$. The offspring group, $C'_{t+1}$, have characteristics dependent on the genetic operators: crossover and mutation. The genetic operators are an implementation decision \cite{FogelDavidB2009}. 

Once the new population has been created, the new population $P_{t+1}$ is created by merging individuals from $C'_{t+1}$ and $P_{t}$. See Algorithm \ref{genetic-algorithm} for detailed pseudocode.

We used the DEAP evolutionary computation framework to create our GA \cite{Gagn2012}. This framework gave us sufficient flexibility when designing our GA. Specifically, it enabled us to persist the data of each generation after every iteration to allow us to verify and analyze our results in real-time.

\begin{algorithm}[t]
\begin{algorithmic}[1]
\State $t=0$
\State initialize $P_{t}$
\State evaluate structures in $P_{t}$
\While {termination condition not satisfied}
\State $t=t+1$
\State select reproduction $C_{t}$ from $P_{t-1}$
\State recombine and mutate structures in $C_{t}$

forming $C'_{t}$
\State evaluate structures in $C'_{t}$
\State select each individual for $P_{t}$ from $C'_{t}$ 

or $P_{t-1}$
\EndWhile
\caption{Genetic algorithm \cite{FogelDavidB2009}}
\label{genetic-algorithm}
\end{algorithmic}
\end{algorithm}

\subsubsection{Parameters for Validation with Observed Data}
\label{ssec:ga_params_valid}

The parameters chosen for the problem explained in Section \ref{sssec:validation} was a population size of $120$, a crossover probability of $50\%$, a mutation probability of $20\%$ and the parameters, $m$ and $c$, as per Equation \ref{eq:problem_formulation}, were given the bounds of $[0.0, 0.004]$ and $[-30, 100]$ respectively. 

The bounds for $m$ and $c$ were calculated to ensure a positive price duration curve, with a maximum price of $\textsterling300$ for $50,000$MW. The population size was chosen to ensure a wide range of solutions could be explored, whilst limiting compute time to ${\sim}1$ day per generation to allow for sufficient verification of the results. The crossover and mutation probabilities were chosen due to suggestions from the DEAP evolutionary computation framework \cite{Gagn2012}.

\subsubsection{Parameters for Long-Term Scenario Analysis}

The parameters chosen for the GA for the problem discussed in Section \ref{sssec:scen-analysis} are displayed here. The population size was $127$, a crossover probability of $50\%$, a mutation probability of $20\%$. The parameters $m_y$, $c_y$ were given the bounds $[0.0, 0.003]$ and $[-30, 50]$ respectively, whilst $\sigma_m$ and $\sigma_c$ were both given the bounds of $[0, 0.001]$. The bounds for $m$ and $c$ were chosen to give an upper bound of \textsterling 200/MWh and lower bound of \textsterling  0/MWh. This is within range of the expected costs of electricity in the UK.

The population size was slightly increased, and the bounds reduced when compared to the parameters for Section \ref{ssec:ga_params_valid}. This was to increase the likelihood of convergence to a global optima, which was more challenging to achieve due to the significantly higher number of parameters.

\section{Results}
\label{sec:results}

Here we present the results of the problem formulation of Sections \ref{sssec:validation} and  \ref{sssec:scen-analysis}. Specifically, we compare the ability of our model to that of BEIS in the context of a historical validation between 2013 and 2018 of the UK electricity market. We also compare our ability to generate scenarios up to 2035 with that of BEIS. 

\subsection{Selecting representative days}
\label{ssec:res_repr_days}

\begin{figure}
\centering
\includegraphics[width=0.49\textwidth]{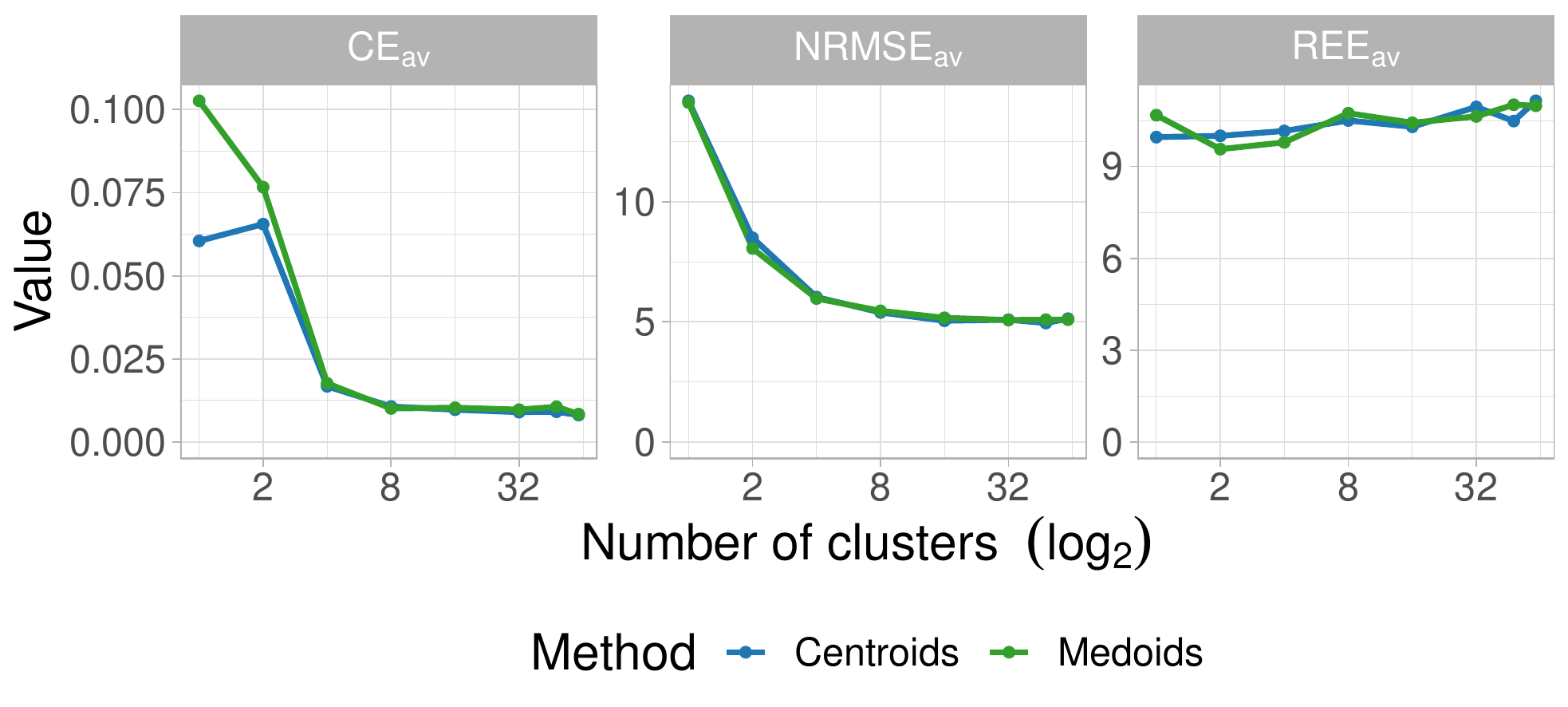}
\caption{Number of clusters compared to error metrics.}
\label{fig:error_metrics_vs_cluster_number}
\end{figure}

Figure \ref{fig:error_metrics_vs_cluster_number} displays the error metrics versus number of clusters. Both $CE_{av}$ and $NRMSE_{av}$ display similar behaviour, namely the error improves significantly from a single cluster to eight clusters for both centroids and medoids. For the number of clusters greater than eight, there are diminishing returns. For $REE_{av}$, however, the error metric is best at a single cluster and gets worse with the number of clusters. It is hypothesised, that $REE_{av}$ worsens over time, as a single cluster in k-means clustering monitors selects most average time-series example. However, as one increases the number of clusters in a k-means algorithm, one trades the $REE_{av}$ for other intricate details surface, such as the distribution of the time-series.

We chose eight clusters as a compromise between the accuracy of the three error metrics and compute time of the simulation. This is because eight was the largest number of clusters that gave us the lowest score for $CE_{av}$, $NRMSE_{av}$ and $REE_{av}$ without significantly increasing compute time. By selecting eight representative days, we reduced the computation time by ${\sim}40\times$. While there was little significant difference between centroid and medoid, we chose to use the medoids because the extreme high and low values would not be lost due to averaging \cite{Hilbers2019}.

\subsection{Validation with Observed Data}

\begin{figure}
\centering
\includegraphics[width=\columnwidth]{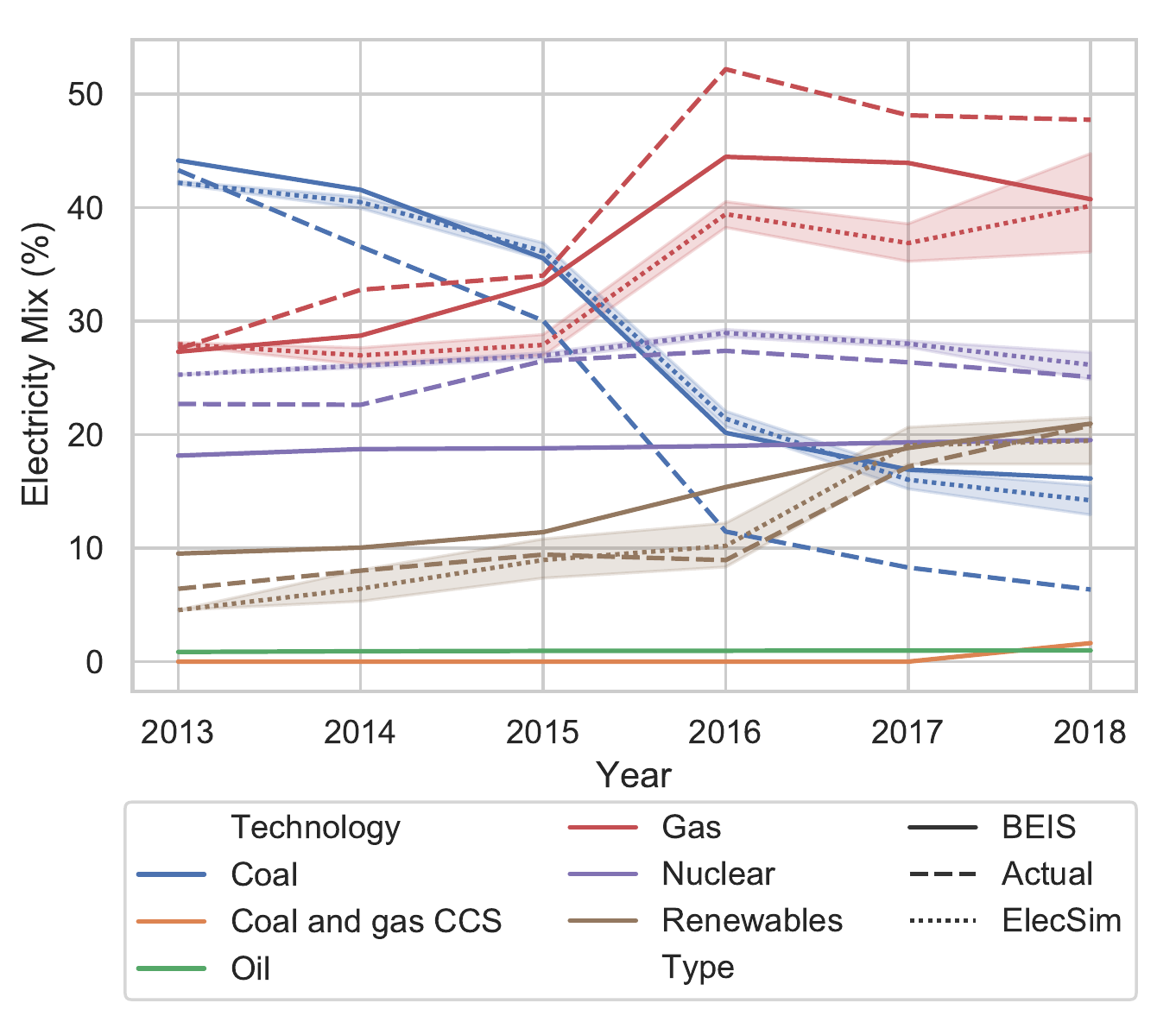}
\caption{Comparison of actual electricity mix vs. ElecSim vs. BEIS projections and taking three coal power plants out of service.}
\label{fig:beis_elecsim_historic_comparison}
\end{figure}

Figure \ref{fig:beis_elecsim_historic_comparison} displays the output of ElecSim under the validation scenario, BEIS' projections and the observed electricity mix between 2013 and 2018, as explained in Sub-Section \ref{sssec:validation}.

The observed electricity mix changed significantly between 2013 and 2018. A continuous decrease in electricity production from coal throughout this period was observed. 2015 and 2016 saw a marked decrease in coal, which can be explained by the retirement of 3 major coal power plants. The decrease in coal between 2013 and 2016 was largely replaced by an increase in gas. After 2016, renewables play an increasingly significant role in the electricity mix and displace gas.

Both ElecSim and BEIS were able to model the fundamental dynamics of this shift from coal to gas as well as the increase in renewables. Both models, however, underestimated the magnitude of the shift from coal to gas. This could be due to unmodelled behaviours such as consumer sentiment towards highly polluting coal plants, a prediction from industry that gas would become more economically attractive in the future or a reaction to The Energy Act 2013 which aimed to close a number of coal power stations over the following two decades \cite{uk_energy_act}.


ElecSim was able to closely model the increase in renewables throughout the period in question, specifically predicting a dramatic increase in 2017. This is in contrast to BEIS, who predicted that an increase in renewable energy would begin in 2016. However, both models were able to predict the proportion of renewables in 2018 accurately. 

ElecSim was able to better model the observed fluctuation in nuclear power in 2016. BEIS, on the other hand, projected a more consistent nuclear energy output. This small increase in nuclear power is likely due to the decrease in coal during that year. BEIS consistently underestimated the share of nuclear power.

We display the error metrics to evaluate our model's five-year projections in Table \ref{table:metrics}. Where MAE is mean absolute squared error, MASE is mean absolute scaled error and RMSE is root mean squared error.

We can improve the projections for all generation types when compared to the naive forecasting approach using ElecSim, as shown by the MASE, where the naive approach is simply predicting the next time-step by using the last known time-step. In this case, the last known time-step is the electricity mix percentage for each generation type in 2013. 

\begin{table}[htb]
    \centering
\begin{filecontents*}{table_data/results/error_metrics.csv}
Technology,MAE,MASE,RMSE
CCGT,9.007,0.701,10.805
Coal,8.739,0.423,10.167
Nuclear,1.69,0.694,2.002
Solar,0.624,0.419,1.019
Wind,1.406,0.361,1.498
\end{filecontents*}
\csvautobooktabular{table_data/results/error_metrics.csv}
    \caption{Error metrics for time series forecast from 2013 to 2018.}
    \label{table:metrics}
    \vspace{-6mm}
\end{table}

\begin{figure}
\centering
\includegraphics[width=0.40\textwidth]{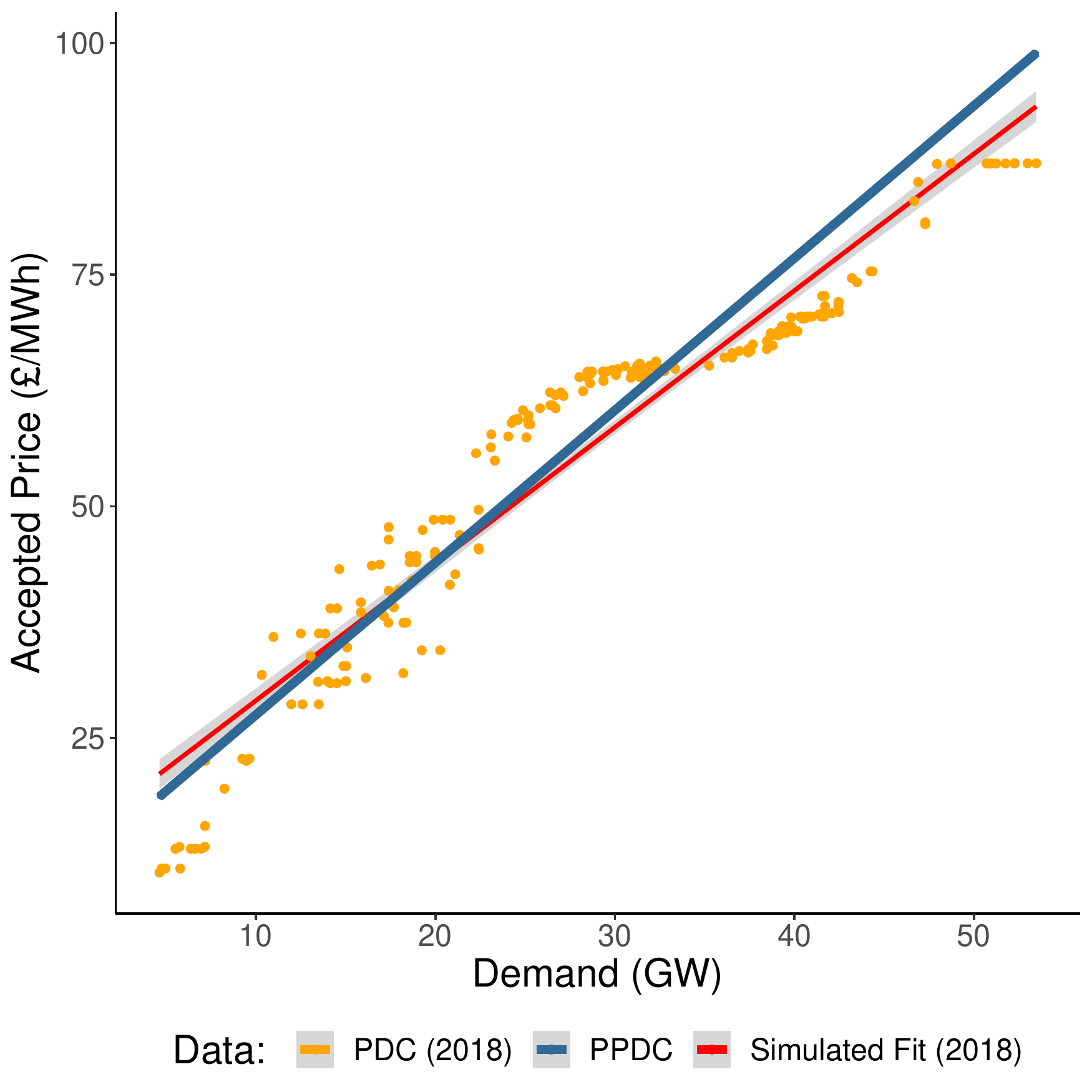}
\caption{Predicted price duration curve for investment for most accurate run against simulated run in 2018.}
\label{fig:best_price_curve}
\end{figure}

Figure \ref{fig:best_price_curve} displays the optimal predicted price duration curve ($PPDC$) found by the GA. This price curve was used by the GenCos to achieve the results shown in Figure \ref{fig:beis_elecsim_historic_comparison}. 

The yellow points show the simulated price duration curve for the first year of the simulation (2018). The red line (Simulated Fit (2018)) is a linear regression that approximates the simulated price duration curve (PDC (2018)). The blue line shows the price duration curve predicted ($PPDC$) by the GenCos to be representative of the expected prices over the lifetime of the plant.


The optimal predicted price duration curve ($PPDC$) closely matches the simulated fit in 2018, shown by Figure \ref{fig:best_price_curve}. However, the $PPDC$ has a slightly higher peak price and lower baseload price. This could be because there is a predicted increase in the number of renewables with a low SRMC. However, due to the intermittency of renewables such as solar and wind, higher peak prices are required to generate in times of low wind and solar irradiance at the earth's surface.

\begin{figure}
\centering
\includegraphics[width=0.45\textwidth]{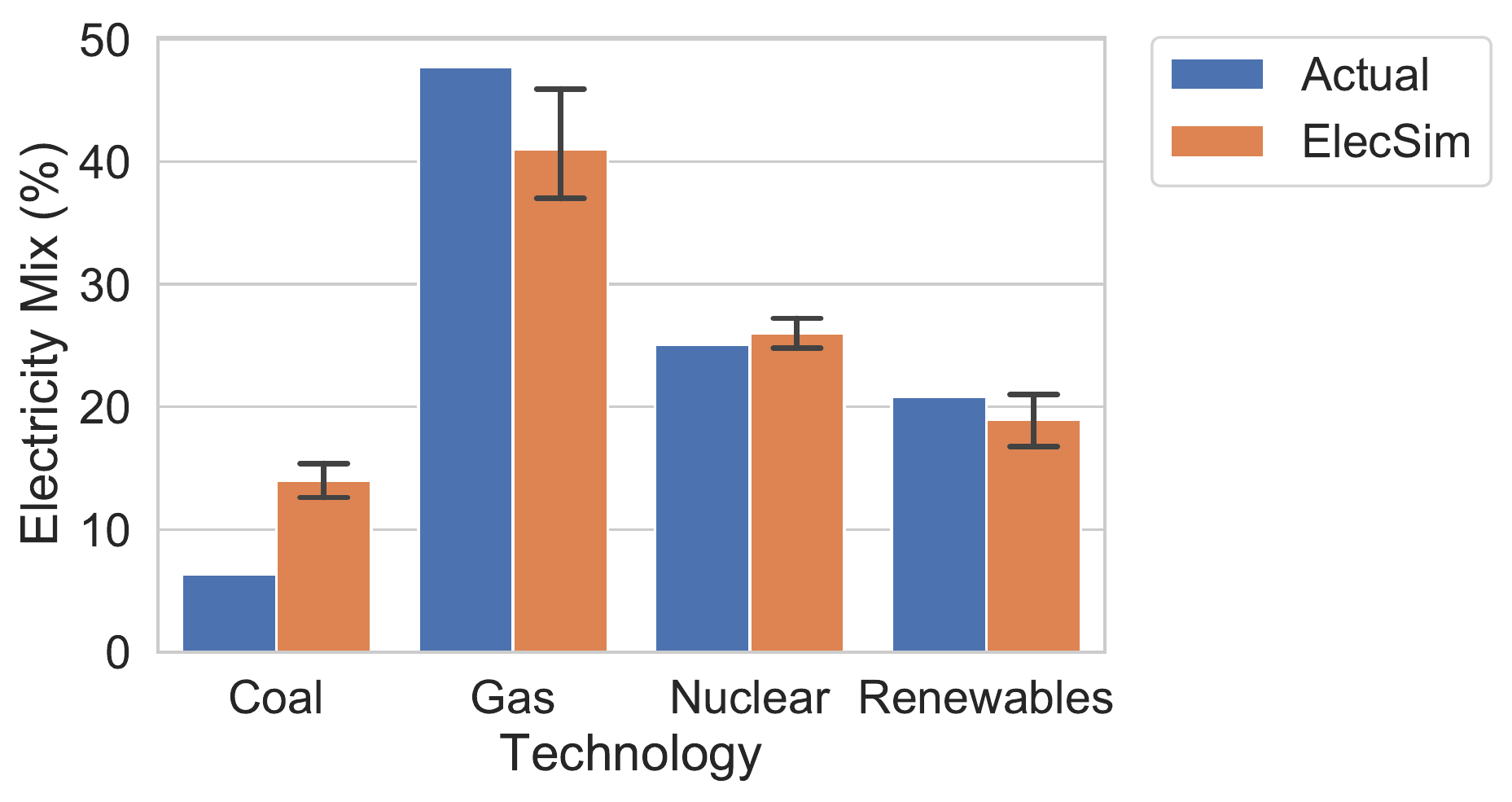}
\caption{Electricity generation mix simulated by ElecSim from 2013 to 2018 compared to observed electricity mix in 2018. The error bars are computed based on a Normal distribution 95\% confidence interval for 40 idential scenario runs.}
\label{fig:uk_validated_results_2018}
\end{figure}

To generate Figure \ref{fig:uk_validated_results_2018}, we ran 40 scenarios with the $PPDC$ to observe the final, simulated electricity mix. The error bars are computed based on a Normal distribution 95\% confidence interval.

ElecSim was able to model the increase in renewables and stability of nuclear energy at this time. ElecSim was also able to model the transition from coal to gas, however, underestimated the magnitude of the transition. This was similar to the projections BEIS made in 2013 as previously discussed.

\subsection{Long-Term Scenario Analysis}

In this section, we discuss the results of the analysis of the BEIS reference scenario explained in Section \ref{sssec:scen-analysis}. Specifically, we created a scenario that mimicked that of BEIS in ElecSim and optimised several parameters using a GA to match this scenario. Through this, we can gain confidence in the underlying dynamics of ElecSim to simulate long-term behaviours. Further, this enables us to verify the likelihood of the scenario by analysing whether the parameters required to make such a scenario are realistic.

\begin{figure}
\centering
\includegraphics[width=0.45\textwidth]{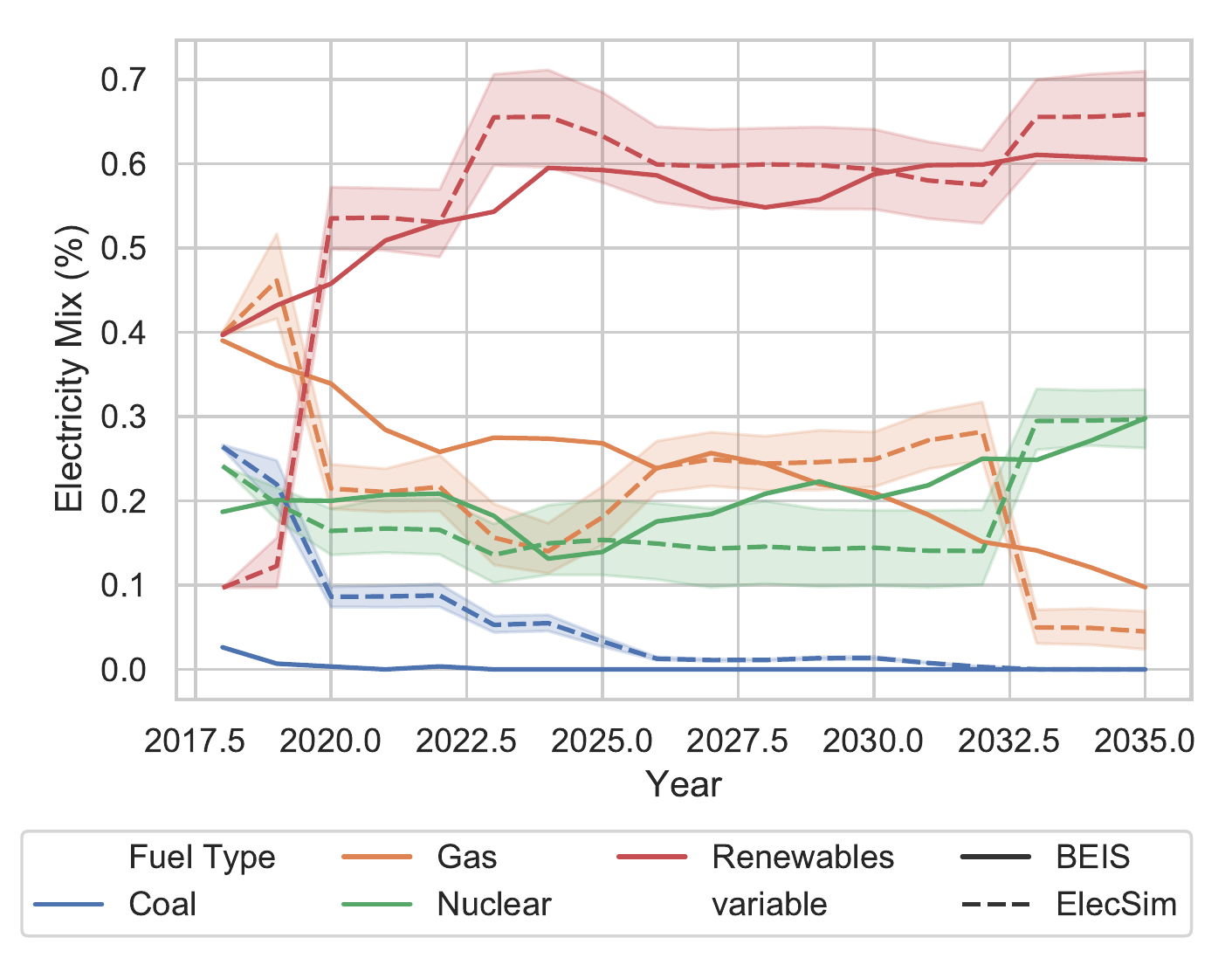}
\caption{Comparison of ElecSim and BEIS' reference scenario from 2018 to 2035.}
\label{fig:forward_scenario_beis_elecsim}
\end{figure}

Figure \ref{fig:forward_scenario_beis_elecsim} displays the electricity mix projected by both ElecSim and BEIS. To generate this image, we ran 60 identical scenarios under the optimal collection of predicted price duration curves, nuclear subsidy and uncertainty in predicted price duration curves, found by running the GA. We ran 60 identical scenarios due to the stochastic sampling of the model, and to generate a probability distribution of each output.

The optimal parameters were chosen by choosing the parameter set with the lowest mean error per electricity generation type and per year throughout the simulation, as shown by Equation \ref{eq:long-term-reward}.

\begin{figure}
\centering
\includegraphics[width=0.45\textwidth, height=0.45\textwidth, keepaspectratio]{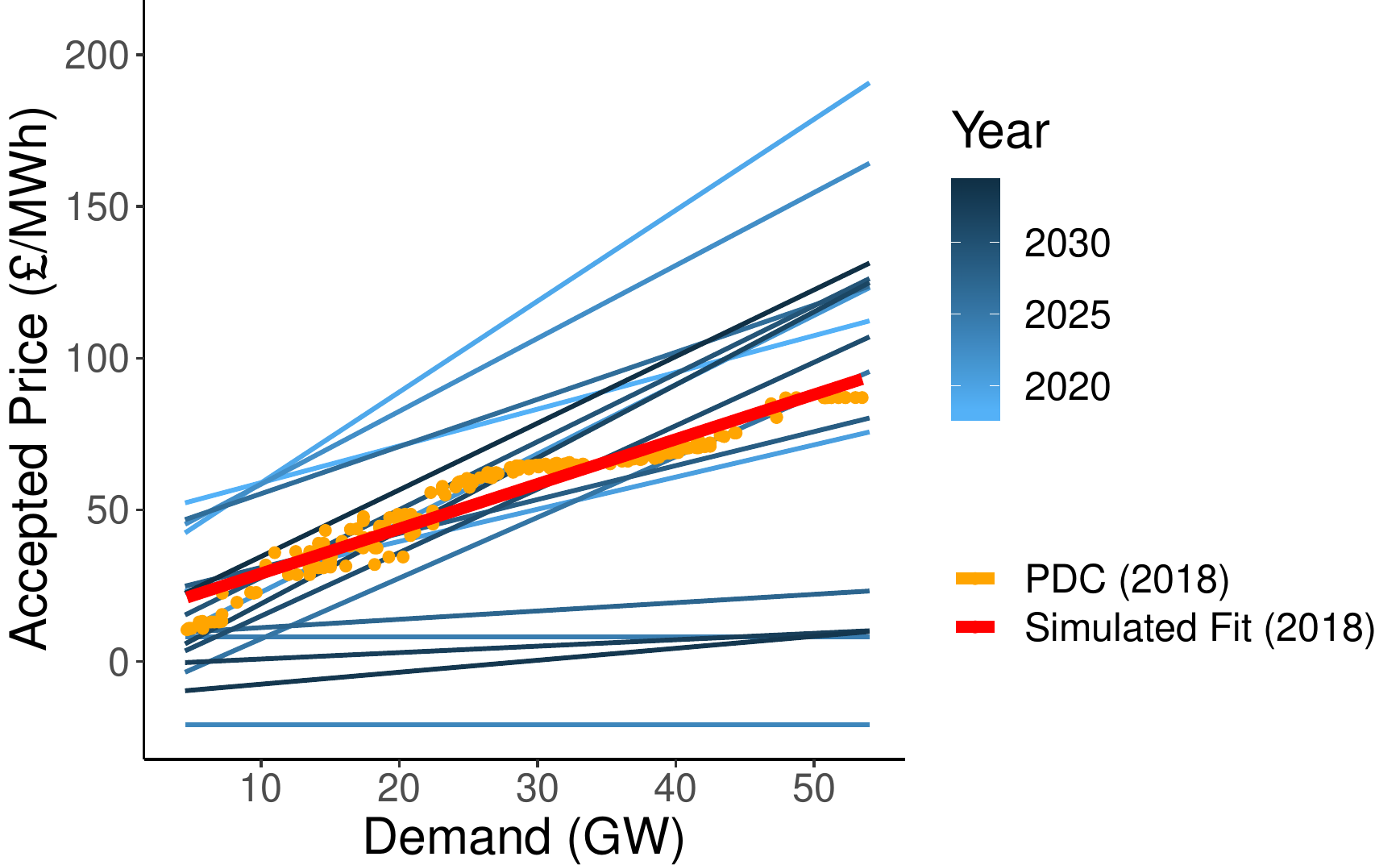}
\caption{Comparison between optimal price duration curves and simulated price duration curve in 2018.}
\label{fig:forward_scenario_best_pdcs}
\vspace{-4mm}
\end{figure}

Figure \ref{fig:forward_scenario_best_pdcs} displays the optimal predicted price duration curves ($PPDC$s) per year of the simulation, shown in blue. These are compared to the price duration curve simulated in 2018, as per Figure \ref{fig:best_price_curve}. The optimal nuclear subsidy, $S_n$, was found to be ${\sim}$\textsterling $120$, the optimal $\sigma_m$ and $\sigma_c$ were found to be $0$ and ${\sim}0.0006$ respectively.

The BEIS scenario demonstrates a progressive increase in nuclear energy from 2025 to 2035, a consistent decrease in electricity produced by natural gas, an increase in renewables and decrease to almost 0\% by 2026 of coal.

ElecSim is largely able to mimic the scenario by BEIS. A large increase in renewables is projected, followed by a decrease in natural gas. A significant difference, however, is the step-change in nuclear power in 2033. This led to an almost equal reduction in natural gas during the same year. In contrast, BEIS project a continuously increasing share of nuclear. 

We argue that the ElecSim projection of nuclear power is more realistic than that of BEIS due to the instantaneous nature of large nuclear power plants coming on-line.

Figure \ref{fig:forward_scenario_best_pdcs} exhibits the price curves required to generate the scenario shown in Figure \ref{fig:forward_scenario_beis_elecsim}. The majority of the price curves are similar to the simulated price duration curve of 2018 (Simulated Fit (2018)). However, there are some price curves which are significantly higher and significantly lower than the predicted price curve of 2018. These cycles in predicted price duration curves may be explained by investment cycles typically exhibited in electricity markets \cite{Gross2007}. 

In this context, investment cycles reflect a boom and bust cycle over long timescales. When electricity supply becomes tight relative to demand, prices rise to create an incentive to invest in new capacity. Price behaviour in competitive markets can lead to periods of several years of low prices \cite{white2005concentrated}. 

As plants retire or demand increases, the market becomes tighter until average prices increase to a level above the threshold for investment in new power generators. At this point, investors may race to bring new plants on-line to make the most out of the higher prices. Once adequate investments have been made, the market returns to a period of low prices and low investment until the next price spike \cite{Gross2007}.

The nuclear subsidy, $S_n$, of ${\sim}$\textsterling $120$ in 2018 prices is high compared to similar subsidies, but this may reflect the difficulty of nuclear competing with renewable technology with a short-run marginal cost that tends to \textsterling $0$.

The low values of $\sigma_m$ and $\sigma_c$ demonstrate that the expectation of prices does not necessarily have to differ significantly between GenCos. This may be due to the fact that GenCos have access to the same market information.

\section{Conclusion}
\label{sec:conclusion}

We have demonstrated that it is possible to use ABMs to simulate liberalised electricity markets. Through validation, we can show that our model, ElecSim, can accurately mimic the observed, real-life scenario in the UK between 2013 and 2018. This provides confidence in the underlying dynamics of ElecSim, especially as we are able to model the fundamental transition between coal and natural gas observed between 2013 and 2018 in the UK.

In addition to this, we were able to compare our long-term scenario to that of the UK Government, Department for Business, Energy \& Industrial strategy. Not only are we able to demonstrate that we can mimic their reference scenario, but we are able to demonstrate a more realistic increase in nuclear power. The parameters that were gained from optimisation show that the BEIS scenario is realistic, however, a high nuclear subsidy may be required.

To improve the accuracy of our model, we used eight representative days of solar irradiance, offshore and onshore wind speed and demand to approximate an entire year. The particular days were chosen using a $k$-means clustering technique, and selecting the medoids. This enabled us to accurately model the daily fluctuations of demand and renewable energy resources and reduce computation time by ${\sim}40\times$. 


In future work, we would like to evaluate further scenarios to provide advice to stakeholders, integrate multi-agent reinforcement learning techniques to better model agents in both investment and bidding strategies as well as model different collections of countries. Further work could be to make predicted price duration curves endogenous to the model. However, this could require scenario analysis by each of the GenCos each time they wanted to make an investment.

In addition to this, a method of dealing with the non-validatable nature of electricity markets, as per the definition of Hodges \textit{et al.} is to vary input parameters over many simulations and look for general trends \cite{Hodges}. This could be achieved using ElecSim through the analysis of a reference case, and a limited set of scenarios which include the most important uncertainties in the model structure, parameters, and data, i.e. alternative scenarios which have both high plausibility and major impacts on the outcomes.

\begin{acks}
This work was supported by the Engineering and Physical Sciences Research Council, Centre for Doctoral Training in Cloud Computing for Big Data [grant number EP/L015358/1].

\end{acks}

\bibliographystyle{ACM-Reference-Format}
\bibliography{custombibtex}

\appendix
\clearpage
\balance
\section{Appendix}

Table \ref{table:modern_plant_costs} shows a sample of modern power plant costs, and Table \ref{table:historic_plant_costs} displays a sample of historic power plant costs. The parameters for both of these tables are explained here.

\textit{Power Plant Parameters.}\label{ssssec:powerplantparameters} Costs form an important element of markets and investment, and publicly available data for power plant costs for individual countries can be scarce. Thus, extrapolation and interpolation can be used to estimate costs for power plants of differing sizes, types and years of construction.

In this work, we initialised costs relevant to the UK. We used the levelized cost of electricity (LCOE) to infer a higher granularity of costs for historical power plants.

The parameters used to initialise the power plants are detailed here: Periods have units of years and costs in \textsterling/MW unless otherwise stated: Efficiency ($\eta$) is defined as the percentage of energy from fuel that is converted into electrical energy (\%). Operating period ($OP$) is the total period in which a power plant is in operation. Pre-development period ($P_D$) and pre-development costs ($P_C$) include the time and costs for pre-licensing, technical and design, as well as costs incurred due to regulatory, licensing and public enquiry. The construction period ($C_D$) and construction costs ($C_C$) are incurred during the development of the plant, excluding network connections. The infrastructure costs ($I_C$) are the costs incurred by the developer in connecting the plant to the electricity or gas grid (\textsterling). Fixed operation \& maintenance costs ($F_C$) are costs incurred in operating the plant that do not vary based on output. Variable operation \& maintenance ($V_C$) costs are incurred in operating the plant that depend on generator output \cite{Ltd2016}.

\begin{table*}[]
	\begin{tabularx}{1.0205\linewidth}{|l|l|c|l|l|l|l|l|l|l|l|l|l|l|}
	\hline
	Type & Capacity & Year & $\eta$ & $OP$ & $P_D$ & $C_D$ & $P_C$ & $C_C$ & $I_C$ & $F_C$ & $V_C$ & $In_C$ & $Con_C$ \\ \hline
	\multirow{3}{*}{CCGT} & 168.0 & 2018/20/25 & 0.34 & 25 & 3 & 3 & 60,000 & 700,000 & 13,600 & 28,200 & 5 & 2,900 & 3,300 \\ \cline{2-14} 
	& 1200.0 & 2018/20/25 & 0.54 & 25 & 3 & 3 & 10,000 & 500,000 & 15,100 & 12,200 & 3 & 2,100 & 3,300 \\ \cline{2-14} 
	& 1471.0 & 2018/20/25 & 0.53 & 25 & 3 & 3 & 10,000 & 500,000 & 15,100 & 11,400 & 3 & 1,900 & 3,300 \\ \hline
	\multirow{5}{*}{Coal} & 552.0 & 2025 & 0.32 & 25 & 6 & 6 & 40,000 & 3,400,000 & 10,000 & 68,200 & 6 & 13,000 & 3,800 \\ \cline{2-14} 
	& 624.0 & 2025 & 0.32 & 25 & 5 & 5 & 70,000 & 4,200,000 & 10,000 & 79,600 & 3 & 19,300 & 3,800 \\ \cline{2-14} 
	& 652.0 & 2025 & 0.3 & 25 & 5 & 5 & 60,000 & 3,900,000 & 10,000 & 65,300 & 5 & 22,700 & 3,800 \\ \cline{2-14} 
	& 734.0 & 2025 & 0.38 & 25 & 5 & 5 & 60,000 & 2,600,000 & 10,000 & 56,400 & 3 & 9,600 & 3,800 \\ \cline{2-14} 
	& 760.0 & 2025 & 0.35 & 25 & 5 & 5 & 40,000 & 2,800,000 & 10,000 & 52,100 & 5 & 14,000 & 3,800 \\ \hline
	\multirow{3}{*}{Hydro} & 0.033 & 2018/20/25 & 1.0 & 35 & 0 & 0 & 0 & 6,300,000 & 0 & 83,300 & 0 & 0 & 0 \\ \cline{2-14} 
	& 1.046 & 2018/20/25 & 1.0 & 35 & 0 & 0 & 0 & 3,300,000 & 400 & 18,200 & 0 & 0 & 0 \\ \cline{2-14} 
	& 11.0 & 2018/20/25 & 1.0 & 41 & 2 & 2 & 60,000 & 3,000,000 & 0 & 45,100 & 6 & 0 & 0 \\ \hline
	Nuclear & 3300.0 & 2025 & 1.0 & 60 & 5 & 8 & 240,000 & 4,100,000 & 11,500 & 72,900 & 5 & 10,000 & 500 \\ \hline
	\multirow{5}{*}{OCGT} & 96.0 & 2018/20/25 & 0.35 & 25 & 2 & 2 & 80,000 & 600,000 & 12,600 & 9,900 & 4 & 2,500 & 2,400 \\ \cline{2-14} 
	& 299.0 & 2018/20/25 & 0.35 & 25 & 2 & 2 & 30,000 & 400,000 & 13,600 & 9,600 & 3 & 1,600 & 2,500 \\ \cline{2-14} 
	& 311.0 & 2018/20/25 & 0.35 & 25 & 2 & 2 & 30,000 & 400,000 & 13,600 & 9,500 & 3 & 1,600 & 2,500 \\ \cline{2-14} 
	& 400.0 & 2018/20/25 & 0.34 & 25 & 2 & 2 & 30,000 & 300,000 & 15,100 & 7,800 & 3 & 1,300 & 2,500 \\ \cline{2-14} 
	& 625.0 & 2018/20/25 & 0.35 & 25 & 2 & 2 & 20,000 & 300,000 & 15,100 & 4,600 & 3 & 1,200 & 2,400 \\ \hline
	\multirow{6}{*}{Offshore} & \multirow{3}{*}{321.0} & 2018 & 0.0 & 23 & 5 & 3 & 60,000 & 2,200,000 & 69,300 & 30,900 & 3 & 1,400 & 33,500 \\ \cline{3-14} 
	&  & 2020 & 0.0 & 23 & 5 & 3 & 60,000 & 2,100,000 & 69,300 & 30,000 & 3 & 1,400 & 32,600 \\ \cline{3-14} 
	&  & 2025 & 0.0 & 23 & 5 & 3 & 60,000 & 1,900,000 & 69,300 & 28,600 & 3 & 1,300 & 31,100 \\ \cline{2-14} 
	& \multirow{3}{*}{844.0} & 2018 & 0.0 & 22 & 5 & 3 & 120,000 & 2,400,000 & 323,000 & 48,600 & 4 & 3,300 & 50,300 \\ \cline{3-14} 
	&  & 2020 & 0.0 & 22 & 5 & 3 & 120,000 & 2,300,000 & 323,000 & 47,300 & 3 & 3,300 & 48,900 \\ \cline{3-14} 
	&  & 2025 & 0.0 & 22 & 5 & 3 & 120,000 & 2,100,000 & 323,000 & 45,400 & 3 & 3,100 & 47,000 \\ \hline
	\multirow{9}{*}{Onshore} & \multirow{3}{*}{0.01} & 2018 & 1.0 & 20 & 0 & 0 & 0 & 3,700,000 & 0 & 29,700 & 0 & 0 & 0 \\ \cline{3-14} 
	&  & 2020 & 1.0 & 20 & 0 & 0 & 0 & 3,600,000 & 0 & 29,600 & 0 & 0 & 0 \\ \cline{3-14} 
	&  & 2025 & 1.0 & 20 & 0 & 0 & 0 & 3,500,000 & 0 & 29,600 & 0 & 0 & 0 \\ \cline{2-14} 
	& \multirow{3}{*}{0.482} & 2018 & 1.0 & 20 & 0 & 0 & 0 & 2,200,000 & 200 & 56,900 & 0 & 0 & 0 \\ \cline{3-14} 
	&  & 2020 & 1.0 & 20 & 0 & 0 & 0 & 2,100,000 & 200 & 56,900 & 0 & 0 & 0 \\ \cline{3-14} 
	&  & 2025 & 1.0 & 20 & 0 & 0 & 0 & 2,000,000 & 200 & 56,700 & 0 & 0 & 0 \\ \cline{2-14} 
	& \multirow{3}{*}{20.0} & 2018 & 0.0 & 24 & 4 & 2 & 110,000 & 1,200,000 & 3,300 & 23,200 & 5 & 1,400 & 3,100 \\ \cline{3-14} 
	&  & 2020 & 0.0 & 24 & 4 & 2 & 110,000 & 1,200,000 & 3,300 & 23,000 & 5 & 1,400 & 3,100 \\ \cline{3-14} 
	&  & 2025 & 0.0 & 24 & 4 & 2 & 110,000 & 1,200,000 & 3,300 & 22,400 & 5 & 1,400 & 3,000 \\ \hline
	\multirow{14}{*}{PV} & \multirow{3}{*}{0.003} & 2018 & 1.0 & 30 & 0 & 0 & 0 & 1,500,000 & 0 & 23,500 & 0 & 0 & 0 \\ \cline{3-14} 
	&  & 2020 & 1.0 & 30 & 0 & 0 & 0 & 1,500,000 & 0 & 23,400 & 0 & 0 & 0 \\ \cline{3-14} 
	&  & 2025 & 1.0 & 30 & 0 & 0 & 0 & 1,400,000 & 0 & 23,200 & 0 & 0 & 0 \\ \cline{2-14} 
	& \multirow{2}{*}{0.455} & 2018 & 1.0 & 30 & 0 & 0 & 0 & 1,000,000 & 200 & 9,400 & 0 & 0 & 0 \\ \cline{3-14} 
	&  & 2025 & 1.0 & 30 & 0 & 0 & 0 & 900,000 & 200 & 9,200 & 0 & 0 & 0 \\ \cline{2-14} 
	& \multirow{3}{*}{1.0} & 2018 & 0.0 & 25 & 1 & 0 & 20,000 & 700,000 & 0 & 6,600 & 3 & 2,600 & 1,300 \\ \cline{3-14} 
	&  & 2020 & 0.0 & 25 & 1 & 0 & 20,000 & 700,000 & 0 & 6,300 & 3 & 2,600 & 1,300 \\ \cline{3-14} 
	&  & 2025 & 0.0 & 25 & 1 & 0 & 20,000 & 600,000 & 0 & 5,900 & 3 & 2,400 & 1,200 \\ \cline{2-14} 
	& \multirow{3}{*}{4.0} & 2018 & 0.0 & 25 & 1 & 0 & 60,000 & 700,000 & 200 & 8,300 & 0 & 1,200 & 1,300 \\ \cline{3-14} 
	&  & 2020 & 0.0 & 25 & 1 & 0 & 60,000 & 700,000 & 200 & 8,000 & 0 & 1,100 & 1,300 \\ \cline{3-14} 
	&  & 2025 & 0.0 & 25 & 1 & 0 & 60,000 & 600,000 & 200 & 7,500 & 0 & 1,100 & 1,200 \\ \cline{2-14} 
	& \multirow{3}{*}{16.0} & 2018 & 0.0 & 25 & 1 & 0 & 70,000 & 700,000 & 400 & 5,600 & 0 & 2,000 & 1,300 \\ \cline{3-14} 
	&  & 2020 & 0.0 & 25 & 1 & 0 & 70,000 & 600,000 & 400 & 5,400 & 0 & 1,900 & 1,300 \\ \cline{3-14} 
	&  & 2025 & 0.0 & 25 & 1 & 0 & 70,000 & 600,000 & 400 & 5,100 & 0 & 1,800 & 1,200 \\ \hline
	Recip. Engine (Diesel) & 20.0 & 2018/20/25 & 0.34 & 15 & 2 & 1 & 10,000 & 300,000 & 2,200 & 10,000 & 2 & 1,000 & -31,900 \\ \hline
	Recip. Engine (Gas) & 20.0 & 2018/20/25 & 0.32 & 15 & 2 & 1 & 10,000 & 300,000 & 3,400 & 10,000 & 2 & 1,000 & -31,900 \\ \hline
		
	\end{tabularx}

	\caption{Modern power plant costs \cite{Department2016}}
	\label{table:modern_plant_costs}
\end{table*}

\begin{table*}[]
	\begin{tabular}{|l|l|l|l|l|l|l|l|l|l|l|l|l|l|}
	\hline
	Type & Capacity & Year & $\eta$ & $OP$ & $P_D$ & $C_D$ & $P_C$ & $C_C$ & $I_C$ & $F_C$ & $V_C$ & $In_C$ & $Con_C$ \\ \hline
	\multirow{12}{*}{CCGT} & \multirow{4}{*}{168.0} & 1980 & 0.34 & 25 & 3 & 3 & 207,345 & 2,419,027 & 46,998 & 97,452 & 22 & 10,021 & 11,403 \\ \cline{3-14} 
	&  & 1990 & 0.34 & 25 & 3 & 3 & 181,208 & 2,114,099 & 41,073 & 85,167 & 13 & 8,758 & 9,966 \\ \cline{3-14} 
	&  & 2000 & 0.34 & 25 & 3 & 3 & 116,407 & 1,358,089 & 26,385 & 54,711 & 10 & 5,626 & 6,402 \\ \cline{3-14} 
	&  & 2010 & 0.34 & 25 & 3 & 3 & 73,530 & 857,857 & 16,666 & 34,559 & 11 & 3,553 & 4,044 \\ \cline{2-14} 
	& \multirow{4}{*}{1200.0} & 1980 & 0.54 & 25 & 3 & 3 & 59,102 & 2,955,138 & 89,245 & 72,105 & 31 & 12,411 & 19,503 \\ \cline{3-14} 
	&  & 1990 & 0.54 & 25 & 3 & 3 & 59,884 & 2,994,246 & 90,426 & 73,059 & 21 & 12,575 & 19,762 \\ \cline{3-14} 
	&  & 2000 & 0.54 & 25 & 3 & 3 & 49,674 & 2,483,747 & 75,009 & 60,603 & 21 & 10,431 & 16,392 \\ \cline{3-14} 
	&  & 2010 & 0.54 & 25 & 3 & 3 & 60,640 & 3,032,008 & 91,566 & 73,981 & 13 & 12,734 & 20,011 \\ \cline{2-14} 
	& \multirow{4}{*}{1471.0} & 1980 & 0.53 & 25 & 3 & 3 & 92,000 & 4,600,023 & 138,920 & 104,880 & 10 & 17,480 & 30,360 \\ \cline{3-14} 
	&  & 1990 & 0.53 & 25 & 3 & 3 & 54,296 & 2,714,817 & 81,987 & 61,897 & 26 & 10,316 & 17,917 \\ \cline{3-14} 
	&  & 2000 & 0.53 & 25 & 3 & 3 & 49,310 & 2,465,515 & 74,458 & 56,213 & 21 & 9,368 & 16,272 \\ \cline{3-14} 
	&  & 2010 & 0.53 & 25 & 3 & 3 & 46,998 & 2,349,947 & 70,968 & 53,578 & 21 & 8,929 & 15,509 \\ \hline
	\multirow{24}{*}{Coal} & \multirow{4}{*}{552.0} & 1980 & 0.32 & 25 & 6 & 6 & 118,041 & 10,033,488 & 29,510 & 201,259 & 22 & 38,363 & 11,213 \\ \cline{3-14} 
	&  & 1990 & 0.32 & 25 & 6 & 6 & 41,766 & 3,550,192 & 10,441 & 71,212 & 2 & 13,574 & 3,967 \\ \cline{3-14} 
	&  & 2000 & 0.32 & 25 & 6 & 6 & 51,429 & 4,371,538 & 12,857 & 87,687 & 3 & 16,714 & 4,885 \\ \cline{3-14} 
	&  & 2010 & 0.32 & 25 & 6 & 6 & 43,411 & 3,689,957 & 10,852 & 74,016 & 10 & 14,108 & 4,124 \\ \cline{2-14} 
	& \multirow{8}{*}{624.0} & 1980 & 0.32 & 25 & 5 & 5 & 183,851 & 11,031,076 & 26,264 & 206,176 & 15 & 41,497 & 9,980 \\ \cline{3-14} 
	&  & 1980 & 0.32 & 25 & 5 & 5 & 188,476 & 11,308,571 & 26,925 & 211,362 & 11 & 42,541 & 10,231 \\ \cline{3-14} 
	&  & 1990 & 0.32 & 25 & 5 & 5 & 62,458 & 3,747,483 & 8,922 & 70,042 & 5 & 14,097 & 3,390 \\ \cline{3-14} 
	&  & 1990 & 0.32 & 25 & 5 & 5 & 65,126 & 3,907,588 & 9,303 & 73,034 & 3 & 14,699 & 3,535 \\ \cline{3-14} 
	&  & 2000 & 0.32 & 25 & 5 & 5 & 80,033 & 4,802,002 & 11,433 & 89,751 & 3 & 18,064 & 4,344 \\ \cline{3-14} 
	&  & 2000 & 0.32 & 25 & 5 & 5 & 80,882 & 4,852,979 & 11,554 & 90,704 & 3 & 18,256 & 4,390 \\ \cline{3-14} 
	&  & 2010 & 0.32 & 25 & 5 & 5 & 84,549 & 5,072,973 & 12,078 & 94,816 & 3 & 19,084 & 4,589 \\ \cline{3-14} 
	&  & 2010 & 0.32 & 25 & 5 & 5 & 81,834 & 4,910,056 & 11,690 & 91,771 & 5 & 18,471 & 4,442 \\ \cline{2-14} 
	& \multirow{4}{*}{652.0} & 1980 & 0.3 & 25 & 5 & 5 & 161,344 & 10,487,387 & 26,890 & 175,596 & 16 & 61,041 & 10,218 \\ \cline{3-14} 
	&  & 1990 & 0.3 & 25 & 5 & 5 & 54,542 & 3,545,235 & 9,090 & 59,359 & 4 & 20,635 & 3,454 \\ \cline{3-14} 
	&  & 2000 & 0.3 & 25 & 5 & 5 & 68,516 & 4,453,581 & 11,419 & 74,568 & 2 & 25,922 & 4,339 \\ \cline{3-14} 
	&  & 2010 & 0.3 & 25 & 5 & 5 & 67,915 & 4,414,497 & 11,319 & 73,914 & 4 & 25,694 & 4,301 \\ \cline{2-14} 
	& \multirow{4}{*}{734.0} & 1980 & 0.38 & 25 & 5 & 5 & 249,766 & 10,823,198 & 41,627 & 234,780 & 16 & 39,962 & 15,818 \\ \cline{3-14} 
	&  & 1990 & 0.38 & 25 & 5 & 5 & 87,920 & 3,809,903 & 14,653 & 82,645 & 7 & 14,067 & 5,568 \\ \cline{3-14} 
	&  & 2000 & 0.38 & 25 & 5 & 5 & 118,072 & 5,116,482 & 19,678 & 110,988 & 5 & 18,891 & 7,477 \\ \cline{3-14} 
	&  & 2010 & 0.38 & 25 & 5 & 5 & 132,370 & 5,736,075 & 22,061 & 124,428 & 5 & 21,179 & 8,383 \\ \cline{2-14} 
	& \multirow{4}{*}{760.0} & 1980 & 0.35 & 25 & 5 & 5 & 160,182 & 11,212,746 & 40,045 & 208,637 & 8 & 56,063 & 15,217 \\ \cline{3-14} 
	&  & 1990 & 0.35 & 25 & 5 & 5 & 55,208 & 3,864,573 & 13,802 & 71,908 & 4 & 19,322 & 5,244 \\ \cline{3-14} 
	&  & 2000 & 0.35 & 25 & 5 & 5 & 65,705 & 4,599,358 & 16,426 & 85,580 & 8 & 22,996 & 6,241 \\ \cline{3-14} 
	&  & 2010 & 0.35 & 25 & 5 & 5 & 77,393 & 5,417,570 & 19,348 & 100,805 & 3 & 27,087 & 7,352 \\ \hline
	\multirow{4}{*}{Nuclear} & \multirow{4}{*}{3300.0} & 1980 & 1.0 & 60 & 5 & 8 & 516,790 & 8,828,507 & 24,762 & 156,975 & 21 & 21,532 & 1,076 \\ \cline{3-14} 
	&  & 1990 & 1.0 & 60 & 5 & 8 & 390,159 & 6,665,224 & 18,695 & 118,510 & 3 & 16,256 & 812 \\ \cline{3-14} 
	&  & 2000 & 1.0 & 60 & 5 & 8 & 378,998 & 6,474,560 & 18,160 & 115,120 & 15 & 15,791 & 789 \\ \cline{3-14} 
	&  & 2010 & 1.0 & 60 & 5 & 8 & 388,457 & 6,636,156 & 18,613 & 117,994 & 13 & 16,185 & 809 \\ \hline
	\multirow{8}{*}{Offshore} & \multirow{4}{*}{321.0} & 1980 & 0.0 & 23 & 5 & 3 & 100,043 & 3,668,254 & 115,550 & 51,522 & 9 & 2,334 & 55,857 \\ \cline{3-14} 
	&  & 1990 & 0.0 & 23 & 5 & 3 & 104,550 & 3,833,513 & 120,755 & 53,843 & 3 & 2,439 & 58,373 \\ \cline{3-14} 
	&  & 2000 & 0.0 & 23 & 5 & 3 & 102,374 & 3,753,742 & 118,242 & 52,723 & 6 & 2,388 & 57,159 \\ \cline{3-14} 
	&  & 2010 & 0.0 & 23 & 5 & 3 & 98,571 & 3,614,292 & 113,850 & 50,764 & 6 & 2,300 & 55,035 \\ \cline{2-14} 
	& \multirow{4}{*}{844.0} & 1980 & 0.0 & 22 & 5 & 3 & 181,469 & 3,629,393 & 488,455 & 73,495 & 8 & 4,990 & 76,066 \\ \cline{3-14} 
	&  & 1990 & 0.0 & 22 & 5 & 3 & 178,822 & 3,576,447 & 481,330 & 72,423 & 10 & 4,917 & 74,956 \\ \cline{3-14} 
	&  & 2000 & 0.0 & 22 & 5 & 3 & 180,212 & 3,604,250 & 485,072 & 72,986 & 9 & 4,955 & 75,539 \\ \cline{3-14} 
	&  & 2010 & 0.0 & 22 & 5 & 3 & 171,372 & 3,427,446 & 461,277 & 69,405 & 11 & 4,712 & 71,833 \\ \hline
	\multirow{4}{*}{Onshore} & \multirow{4}{*}{20.0} & 1980 & 0.0 & 24 & 4 & 2 & 374,087 & 4,080,950 & 11,222 & 78,898 & 26 & 4,761 & 10,542 \\ \cline{3-14} 
	&  & 1990 & 0.0 & 24 & 4 & 2 & 411,234 & 4,486,197 & 12,337 & 86,733 & 10 & 5,233 & 11,589 \\ \cline{3-14} 
	&  & 2000 & 0.0 & 24 & 4 & 2 & 230,491 & 2,514,457 & 6,914 & 48,612 & 5 & 2,933 & 6,495 \\ \cline{3-14} 
	&  & 2010 & 0.0 & 24 & 4 & 2 & 143,450 & 1,564,915 & 4,303 & 30,255 & 7 & 1,825 & 4,042 \\ \hline
	\multirow{4}{*}{PV} & \multirow{4}{*}{16.0} & 1980 & 0.0 & 25 & 1 & 0 & 399,799 & 3,997,991 & 2,284 & 31,983 & 0 & 11,422 & 7,424 \\ \cline{3-14} 
	&  & 1990 & 0.0 & 25 & 1 & 0 & 399,799 & 3,997,991 & 2,284 & 31,983 & 0 & 11,422 & 7,424 \\ \cline{3-14} 
	&  & 2000 & 0.0 & 25 & 1 & 0 & 399,799 & 3,997,991 & 2,284 & 31,983 & 0 & 11,422 & 7,424 \\ \cline{3-14} 
	&  & 2010 & 0.0 & 25 & 1 & 0 & 399,799 & 3,997,991 & 2,284 & 31,983 & 0 & 11,422 & 7,424 \\ \hline
	\end{tabular}
	\caption{Sample of historic power plant costs \cite{IRENA2018,IEA2015,OWPB2016}} 
	\label{table:historic_plant_costs}

\end{table*}

\clearpage
\end{document}